\documentclass[aps,pre,twocolumn,superscriptaddress,showpacs]{revtex4-1}

\usepackage{amsmath,amssymb,bm}
\usepackage{graphics}
\usepackage{graphicx}
\usepackage{comment}
\usepackage{mathtools}
\begin{document}
\title{
Inverting the Achlioptas rule for explosive percolation 
}

\author{R.~A.~da~Costa}
\affiliation{Departamento de F{\'\i}sica da Universidade de Aveiro $\&$ I3N, Campus Universit\'ario de Santiago, 3810-193 Aveiro, Portugal}
\author{S.~N. Dorogovtsev}
\affiliation{Departamento de F{\'\i}sica da Universidade de Aveiro $\&$ I3N, Campus Universit\'ario de Santiago, 3810-193 Aveiro, Portugal}
\affiliation{A.~F. Ioffe Physico-Technical Institute, 194021 St. Petersburg, Russia}
\author{A.~V. Goltsev}
\affiliation{Departamento de F{\'\i}sica da Universidade de Aveiro $\&$ I3N, Campus Universit\'ario de Santiago, 3810-193 Aveiro, Portugal}
\affiliation{A.~F. Ioffe Physico-Technical Institute, 194021 St. Petersburg, Russia}
\author{J.~F.~F. Mendes}
\affiliation{Departamento de F{\'\i}sica da Universidade de Aveiro $\&$ I3N, Campus Universit\'ario de Santiago, 3810-193 Aveiro, Portugal}
\begin{abstract} 
In the usual Achlioptas processes the smallest clusters of a few randomly chosen ones are selected to merge together at each step. 
The resulting aggregation process leads to the delayed birth of a giant cluster and the so-called explosive percolation transition showing a set of anomalous features. 
We explore a process with the opposite selection rule, in which the biggest clusters of the randomly chosen ones merge together. 
We develop a theory of this kind of percolation based on the Smoluchowski equation, find the percolation threshold, and describe the scaling properties of this continuous transition, 
namely, the critical exponents and amplitudes, and scaling functions. We show that, qualitatively, this transition is similar to the ordinary percolation one,
though occurring in less connected systems.

\end{abstract}
\maketitle



Aggregation processes based on progressive merging together the smallest clusters of a few randomly chosen (Achlioptas rule) have attracted much 
attention in the last years \cite{achlioptas2009explosive,spencer2007birth,ziff2009explosive,da2010explosive,d2010local,nagler2011impact,araujo2011tricritical,grassberger2011explosive,da2014critical,fortunato2011explosive,lee2011continuity,riordan2011explosive,riordan2012achlioptas,cho2013avoiding,araujo2014recent,bastas2014explosive}. These specific processes and models show a number of unusual features \cite{da2014solution} distinguishing them sharply from ordinary aggregation processes and standard percolation, in which random clusters merge together with probability proportional to their sizes \cite{stauffer1991introduction,stauffer1979scaling,dorogovtsev2008critical}. Apart from the delayed percolation phase transition, which is continuous, as we 
found for a wide range of systems \cite{da2010explosive,da2014critical} and which was proven mathematically \cite{riordan2011explosive,riordan2012achlioptas}, these models demonstrate a uniquely small exponent $\beta$ of the percolation cluster and unusual scaling functions. The smallness of $\beta$ makes the transition so ``sharp'' that it is difficult to distinguish it from discontinuous in simulations, which resulted in the term ``explosive percolation'' \cite{achlioptas2009explosive}. 
A few real-world applications of these processes were identified \cite{rozenfeld2010explosive,kim2010explosive,pan2011using}. 
The Achlioptas processes, generalizing percolation, constitute a wide class 
including the processes generated by the original ``product rule'' (two clusters with the smallest product of sizes are selected) \cite{achlioptas2009explosive}, the sum rule (clusters with the smallest sum of sizes are selected), the rule selecting the smallest clusters \cite{da2010explosive}, and many others, of which only a small number were explored. 
The problem is how far from the standard percolation scenario can these diverse rules and their variations lead? How easy can one deviate from the typical percolation behavior by exploiting the ``power of choice'' \cite{d2007power} in these processes?
Notably, in these rules another kind of optimization can be considered, 
namely, selecting not the smallest but the largest clusters. 
In particular, the 
question is: what will happen if we invert the Achlioptas rule, 
that is, at each step merge together the two largest clusters of a few randomly selected ones \cite{tanaka2013network,giazitzidis2014variation}? 

In the present article we answer to this question by considering a representative set of 
processes based on the inverse Achlioptas rule, for which we derive 
the Smoluchowski equation. By solving these equations numerically and analytically we find that this rule results in a percolation transition taking place at an earlier stage of the process, 
but with the same set of critical exponents as in ordinary percolation. We calculate the critical amplitudes for the relative size $S$ of the percolation cluster and for the size distribution of finite clusters and obtain the scaling functions. 

The paper is organized as follows. In Sec.~\ref{s1} we introduce the model and describe the evolution equations. In Sec.~\ref{s2} we obtain the critical singularity of the percolation cluster $S$ by using the generating function technique.    
In Sec.~\ref{s3} we find the scaling functions and the critical exponent $\tau$, $P(s,t_c) \sim s^{1-\tau}$.   
Section~\ref{s4} describes the order parameter and generalized susceptibility for these phase transitions. 
Finally in Sec.~\ref{s5} we obtain analytical estimates for the transition point and critical amplitude. Table~\ref{t1} demonstrates a good agreement between the results of the numerical solution of evolution equations and these estimates.


\section{The model}
\label{s1}

We consider the following model incorporating the inverse Achlioptas rule and convenient for treatment. We start from $N$ isolated nodes. At each time step a new link connecting two nodes is added to the
network as follows. 
At each step sample two times: (i)
choose $m\geq 1$ nodes uniformly at random and compare
the clusters to which these nodes belong; select the node
within the largest of these clusters; (ii) similarly choose
the second sample of $m$ nodes and, again, as in (i),
select the node belonging to the largest of the $m$ clusters;
(iii) add a link between the two selected nodes thus
merging the two largest clusters. 
Note that the only difference from our previous works is that instead of selecting the two smallest clusters for merging \cite{da2010explosive,da2014solution,da2015solution}, here we select the two largest. The probability distribution $P(s,t)$, 
i.e., the probability that a uniformly randomly chosen node belongs to a cluster of size $s$ at time $t$,  gives the complete description of the evolution of this system. Time $t$ is the ratio of the number of steps (links) and the number of nodes. 
We emphasize that,  
instead of the product rule \cite{achlioptas2009explosive},  
we select for merging the largest cluster from each of the two sets of $m$ clusters,  
which makes our problem 
treatable analytically.


For infinite $N$, this aggregation process is described by the following evolution equation:
\begin{equation}
\frac{\partial P(s,t)}{\partial t}
= s \sum_{u=1}^{s-1} Q(u,t)Q(s-u,t) - 2 sQ(s,t)
,
\label{11}
\end{equation}
which is  the Smoluchowski equation for this process \cite{smoluchowski1916brownsche,krapivsky2010kinetic}. 
Here $Q(s, t)$ is the probability that a cluster selected to
merge is of size $s$. The distribution $Q(s, t)$ is expressed in terms of $P(s,t)$ as 
\begin{eqnarray}
Q(s)&& = \left[\sum_{u=1}^{s}P(u)\right]^{m} - \left[\sum_{u=1}^{s-1}P(u)\right]^{m},
\nonumber
\\
&&\cong m P(s)\left[ \sum_{u<s} P(u) \right]^{m-1}
\label{12}
.
\end{eqnarray}
Notice the normalization conditions for these distributions, $\sum_s P(s)=1{-}S$ and $\sum_s Q(s)=(1{-}S)^m$, where $S$ is the relative size of the percolation cluster.
For large $s$ we have 
\begin{equation}
Q(s)\cong m P(s)\left( 1-S \right)^{m-1}
\label{13}
,
\end{equation}
both above and below $t_c$.
Equation~(\ref{11}) together with relation (\ref{12}) describe the process exactly in the full range of $t$, from $0$ to infinity.


\section{Percolation cluster size}
\label{s2}

Let us define the generating functions 
\begin{equation}
 \rho(z,t) = \sum_s z^s P(s,t) 
 \label{14}
\end{equation}
and
\begin{equation}
\sigma(z,t) = \sum_s z^s Q(s,t) 
\label{14_1}
 .
\end{equation}
We multiply both sides of Eq.~(\ref{11}) by $z^s$ and sum over $s$, which gives 
\begin{equation}
\frac{\partial \rho(z,t)}{\partial t} = 2 \left[\sigma(z,t) - 1\right]\frac{\partial \sigma(z,t)}{\partial \ln z}
\label{15}
 .
\end{equation}
Using Eq.~(\ref{13}), 
we find the relation between the functions $\sigma(z)$ and $\rho(z)$ for $z$ close to $1$, 
\begin{eqnarray}
\sigma(z)\ 
&& 
=\sum_{s}  Q(s) + \sum_{s}  Q(s) (z^s-1)
\nonumber
\\[5pt]
&&  \cong (1-S)^m +  m( 1-S)^{m-1} \sum_{s}  P(s) (z^s-1)
\nonumber
\\[5pt]
&&= m (1-S)^{m-1}\rho(z)-(m-1)(1-S)^m
,
\label{e16}
\end{eqnarray}
and so 
\begin{eqnarray}
\frac{\partial \rho}{\partial t} &&\cong 2 \left[ m(1-S)^{m-1}\rho-(m-1)(1-S)^m -1 \right] 
\nonumber
\\[5pt]
&& \times  m(1-S)^{m-1}\frac{\partial \rho}{\partial \ln z}
.
\label{17}
\end{eqnarray}
By applying hodograph transformation \cite{krapivsky2010kinetic} to this partial differential equation we get the ordinary differential equation 
\begin{eqnarray}
\frac{d \ln z}{d t}\bigg|_{\rho}
&&\cong  2 m(1-S)^{m-1} 
\nonumber
\\[5pt]
&& \times  \left[ (m-1)(1-S)^m +1 -m(1-S)^{m-1}\rho \right] 
,
\label{18}
\end{eqnarray}
which leads to
\begin{eqnarray}
\ln z \cong && \ 2 m (m{-}1) \! \int_{t_c}^{t} \! dt\, (1{-}S)^{2m-1} +2m\! \int_{t_c}^{t}\! dt\, (1{-}S)^{m-1}  
\nonumber
\\[5pt]
&&-2 m^2 \rho\! \int_{t_c}^{t} \! dt\, (1-S)^{2m-2}+ g(\rho)
,
\label{19}
\end{eqnarray}
where the function $g(\rho)$ is the initial condition for Eq.~(\ref{18}), which can be found from the critical distribution $P(s,t_c)\cong a_0 s^{1-\tau}$. 
At $t=t_c$, proceeding as in \cite{da2010explosive,da2014solution,da2015solution} for $z$ close to $1$, we find the singularity 
\begin{equation}
1-\rho(z,t_c) \cong -a_0 \Gamma(2-\tilde{\tau})\left(-\ln z\right)^{\tilde{\tau}-2}
.
\label{111}
\end{equation}
Inverting the function $\rho(z,t_c)$ in this equation we obtain 
\begin{equation*}
g(\rho)= - \left[\frac{-(1-\rho)}{a_0\Gamma(2{-}\tilde{\tau})}\right]^{1/(\tilde{\tau}-2)}
,
\end{equation*}
which we substitute into Eq.~(\ref{19}). Finally, we set $z=1$ in the resulting equation and obtain
\begin{eqnarray}
&&0 \cong   2m(m{-}1)\int_{t_c}^{t} dt\,(1-S)^{2m-1} +  2m  \int_{t_c}^{t} dt\, (1-S)^{m-1}  
\nonumber
\\[5pt]
&&
- 2 m^2 (1-S)\!\! \int_{t_c}^{t} \!\!\!dt\, (1{-}S)^{2m-2}  - \left[\frac{-S}{a_0\Gamma(2{-}\tilde{\tau})}\right]^{1/(\tilde{\tau}-2)}
\!\!\!\!\!.
\label{112}
\end{eqnarray}
From this equation 
we find
 the critical singularity of $S$,
\begin{equation}
 S\cong \left[-a_0\Gamma(2-\tau)\right]^{1/(3-\tau)}  [2m^2 (t-t_c)]^{(\tau-2)/(3-\tau)}
.
\label{114}
\end{equation}


\section{Scaling properties 
}
\label{s3}

In this section we find the scaling form of $P(s,t)$ near $t_c$ 
using the approach of our previous works~\cite{da2014solution,da2015solution}. 
The form of the scaling function is determined by the 
critical exponent $\tau$. 
The scaling function must decay faster than any power law, and must take only positive values. 
We show that these conditions are satisfied only for $\tau=5/2$, which enables us to find the scaling function for each $m$.

\begin{figure*}[t]
\begin{center}
\includegraphics[scale=0.45]{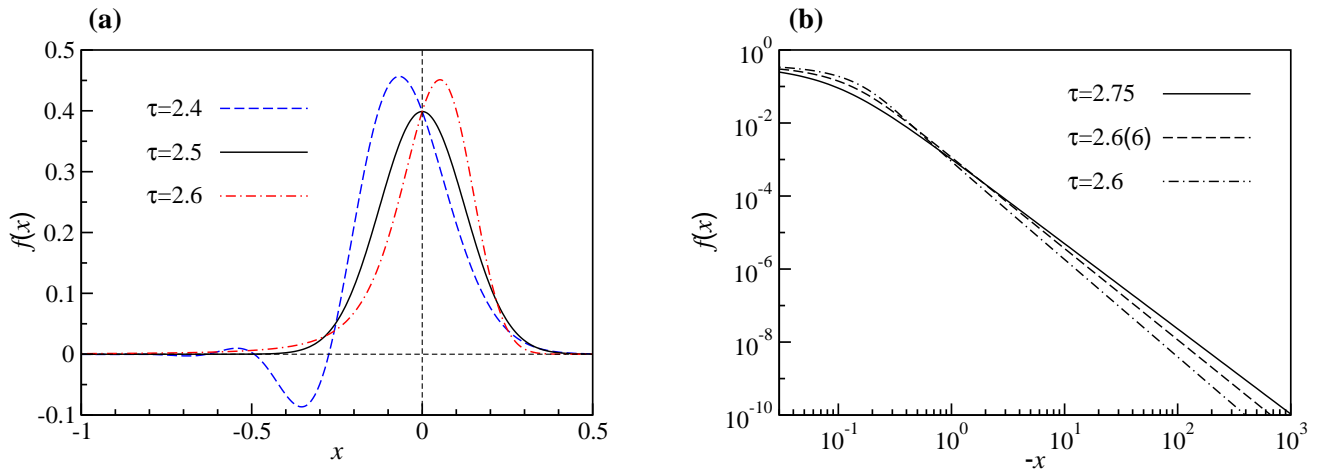}
\ \ \ \ \ \ \ \ \ 
\raisebox{-4pt}{\includegraphics[scale=0.45]{scal_func_f2_v2.eps}}
\end{center}
\caption{Functions $f(x)$ of Eq.~(\ref{120}) for $a_0=1/\sqrt{2 \pi}$ and different values of $\tau$. (a) 
Only 
$\tau=5/2$ provides a rapidly decaying positive function. 
(b) In the phase below $t_c$ (i.e., $x<0$), 
the values $\tau>5/2$ result in functions decaying in a power-law way. 
}
\label{f1}
\end{figure*}

Let us 
obtain the Taylor expansion of $P(s,t)$, 
\begin{equation}
P(s,t)=A_0(s) + A_1(s)(t-t_c)+A_2(s)(t-t_c)^2+...
\label{115}
\end{equation}
in terms of 
$\tau$, by sequentially differentiating the evolution equation~(\ref{11}) and the relation~(\ref{13}) 
at $t=t_c$ with respect to $t$.  
Recall that 
the critical distribution $Q(s,t_c)\cong mP(s,t_c)$ for large $s$, and so the derivatives are 
\begin{equation}
\frac{\partial^n Q(s,t)}{\partial t^n}\bigg|_{t=t_c} \cong  m \frac{\partial^n P(s,t)}{\partial t^n}\bigg|_{t=t_c} 
.
\label{116}
\end{equation}
Differentiating both sides of Eq.~(\ref{11}) $n-1$ times and replacing the right-hand side with Eq.~(\ref{116}) we find the asymptotics of the coefficient $A_n(s)\equiv \partial^{(n)}_t P(s,t)|_{t_c}/n!$. 
Due to 
Eq.~(\ref{116}) the equations for $A_{n}$ for each $m\geq1$ differ only by the factor $m^{2n}$ on the right-hand side. So the asymptotics $A_n(s)$ is
\begin{equation}
A_n(s)\cong a_n m^{2n} s^{1-\tau+n(3-\tau)}
,
\label{117}
\end{equation}
with the prefactor $a_n$  
that we have calculated in~\cite{da2015solution} for ordinary percolation ($m=1$), 
\begin{equation}
a_n= \frac{ 2^n \left[ a_0 \Gamma(2-\tau)\right]^{n+1}}{ (n+1)! \,\Gamma[(n+1)(2-\tau)]} 
.
\label{118}
\end{equation}
The scaling form of the distribution $P(s,t)$ is 
\begin{eqnarray}
P(s,t)
&\cong & s^{1-\tau} \sum_n a_n [m^2  s^{3-\tau}(t-t_c)]^{n}
\nonumber
\\[5pt]
&\cong & s^{1-\tau} f[ s^{3-\tau}(t-t_c)]
,
\label{119}
\end{eqnarray}
where $f(x)$ is the series
\begin{equation}
f(x)\cong   a_0 \Gamma(2-\tau) \sum_{n=0}^{\infty}  \frac{ \left[2 a_0 m^2 \Gamma(2-\tau)x\right]^{n} }{ (n+1)! \,\Gamma[(n+1)(2-\tau)]} 
.
\label{120}
\end{equation}
The parameter $m$ appears only as a factor of $x$. 
The function $f(x)$ depends essentially on $\tau$. For $x \gg 1$ this function approaches $0$ exponentially, staying positive, for any $2<\tau<3$. 
In the phase $t<t_c$, i.e., $x<0$, only one value of the exponent $\tau$ results in a scaling function $f(x)$ with the proper decay to $0$ as $x$ approaches $-\infty$. 
For $\tau<5/2$, the function $f(x)$, Eq.~(\ref{120}), oscillates around $0$ in the region $x<0$, see Fig.~\ref{f1}(a). Since the scaling function cannot take negative values, we exclude the range $\tau<5/2$  from the possible values of $\tau$. 
For $\tau>5/2$, the function 
$f(x)$ stays positive but approaches $0$ as a power-law as $x\to-\infty$, see Fig.~\ref{f1}(b). The scaling function 
 must decay 
 more rapidly than any power law for  $x\to \pm\infty$, and so we also exclude the range $\tau>5/2$. 

\begin{figure*}[t]
\begin{center}
\includegraphics[scale=0.45]{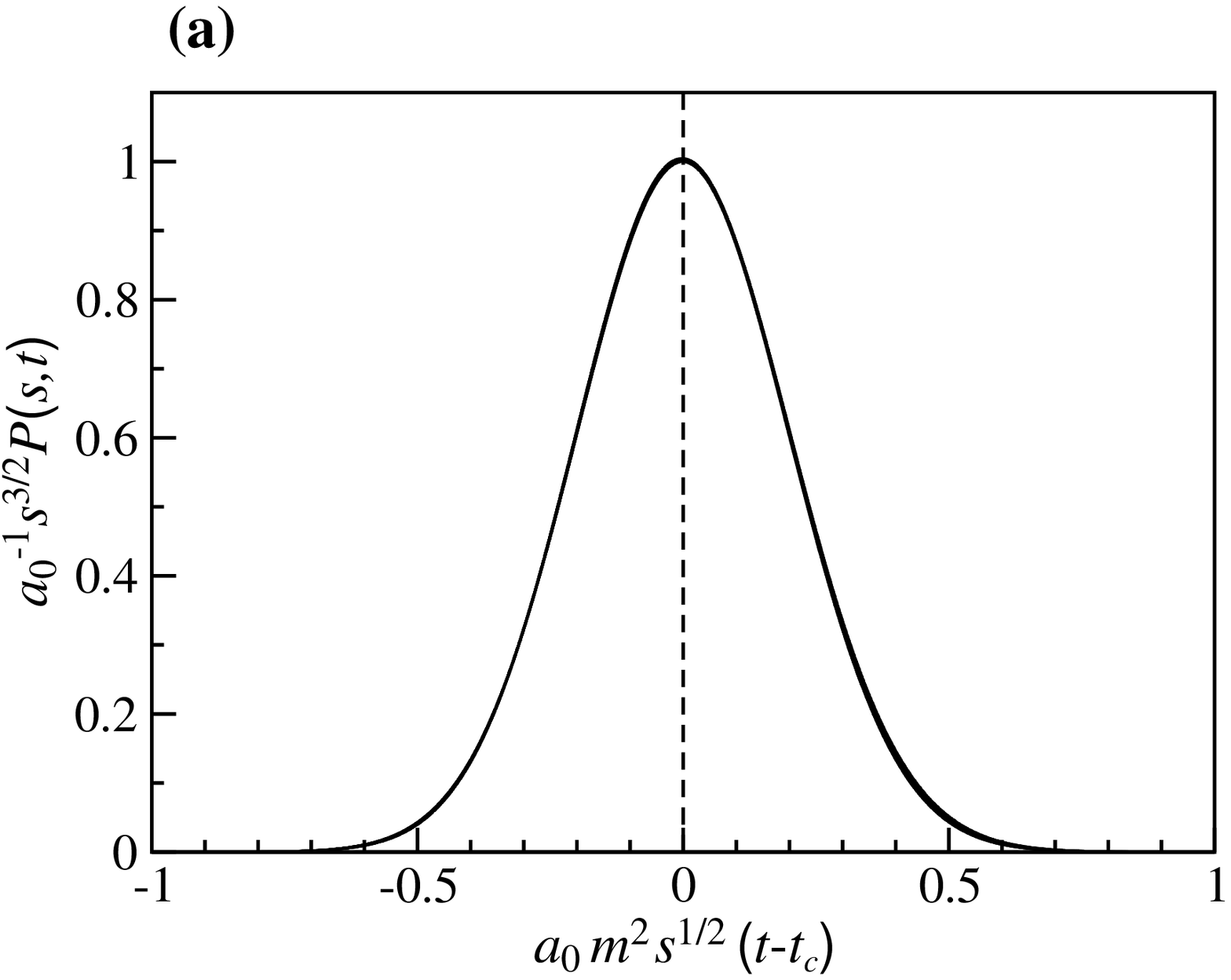}
\ \ \ \ \ \ \ \ \ 
\raisebox{4.5pt}{\includegraphics[scale=0.45]{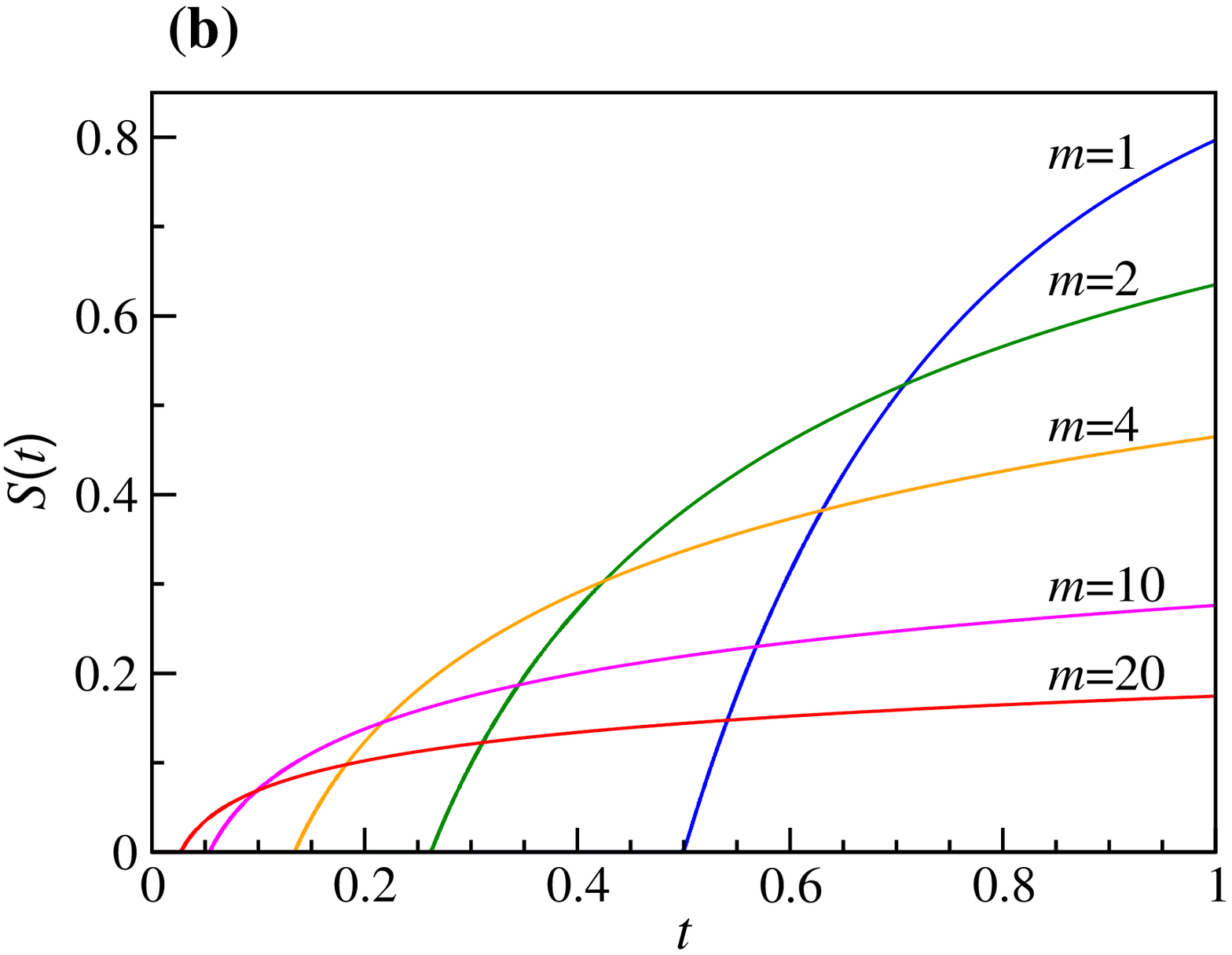}}
\end{center}
\caption{Results of numerical solution of evolution equation~(\ref{11}) for $s\leq10^5$ and different $m$. (a) Evolution of relative size of the percolation cluster $S(t)$. For each $m$, the percolation cluster emerges with an exponent $\beta=1$ at the critical point $t_c(m)$ given in Table~\ref{t1}. 
(b) Rescaled distribution $P(s,t)$ shown in terms of $a_0 m^2 s^{1/2}(t-t_c)$. All curves  $P(s,t,m)$ 
for $14$ combinations of $s=10^4$, $10^5$, and $m=1$, $2$, $3$, $4$, $5$, $10$, $20$, collapse into one, $\exp[-4\pi a_0^2 m^4 s(t-t_c)^2]$, see Eq.~(\ref{121}). 
%
}
\label{f2}
\end{figure*}

At 
$\tau=5/2$ the function $f(x)$ 
takes the form
\begin{eqnarray}
f(x) &\cong&    a_0 \sum_{n=0}^{\infty}  \frac{ \left[-4\pi \left(a_0 m^2 x\right)^2\right]^{n} }{ n!} 
\nonumber
\\[5pt]
&=&
a_0 \exp\left[ -4\pi\left(a_0 m^2 x\right)^2 \right]
,
\label{121}
\end{eqnarray}
that is, decays exponentially with $x^2$ on the both sides of the transition. 
Thus $\tau=5/2$  
for all $m\geq 1$, and so Eq.~(\ref{121}) gives the scaling function of this transition. 
Figure ~\ref{f2}(a) shows that 
near $t_c$ at large $s$ the numerical solution $P(s,t)$ of Eqs. (\ref{11}) and (\ref{12}) agrees completely with the scaling functions (\ref{121}). 
Inserting $\tau=5/2$ into Eq.~(\ref{114}) we find 
\begin{equation}
S\cong 8 \pi m^2 a_0^2 (t-t_c)
\label{123}
\end{equation}
near $t_c$. 
Figure~\ref{f2}(b) presents the evolution of the relative size of the percolation cluster for each $m$, which we found numerically from Eqs.~(\ref{11}) and (\ref{12}), $S(t) = 1-\sum_{s}P(s,t)$. 
The curves $S(t,m)$ in Fig.~\ref{f2}(b) intersect with each other, and so $S(t)$ grows slower for larger $m$ above $t_c$. 
Recall that in our model the percolation cluster can be selected more than once at the same step. This corresponds to adding a new link between two nodes in the percolation cluster, which does not changes cluster sizes and happens with probability $[1-(1-S)^m]^2$. 
This probability 
grows rapidly with $m$  
effectively delaying the aggregation process above $t_c$ compared with $m=1$.
If we forbid the same cluster 
from being selected more than once at each step, the delay effect disappears and a larger $m$ results in a faster growth of $S$, which approaches $S\cong t$ when $m \to \infty$.
In Table~\ref{t1} we show precise results for $\tau$, $t_c$, and $a_0$, which are computed from $P(s\leq10^5,t)$ using the method of Ref.~\cite{da2014critical}. 
The numerical results for $\tau$ agree with the exact result $\tau=5/2$. 
Above the upper critical dimension, which is the case for our models, all critical exponents can be expressed in terms of a single one. So we arrive at the same 
critical exponents as in ordinary percolation.

\begin{table*}[t]
\begin{center}
\begin{tabular}{ c | c | c | c | c | c | c }
\hline
&&&&&&
\\[-11pt]
  $m$ & $\tau$ & $t_c$ &\ \ \ \  $t_c^e=1/(2m)$ \ \ \ \ & $a_0$ &\ \ \ \  $a_0^e=1/(\sqrt{2\pi}\,m)$ \ \ \ \ & $B$ \\[4pt]
\hline
&&&&&&
\\[-11pt]
\ \ \ \ 1  \ \ \ \ &\ \ \ \  $2.5000000(2)$\ \ \ \ &\ \ \ \ $0.500000000(1)$\ \ \ \ &  0.5   &\ \ \ \ $0.398942(3)$ \ \ \ \ &\ \ \ \  0.3989422... \ \ \ \ & \ \ \ \ 4.0000(1)  \ \ \ \   \\[2pt]
  2  & $2.49998(5)$   &   $0.2624198(1)$  & 0.25 & $0.1755(1)$   &  0.1994711...  &  3.096(4)    \\[2pt]
  3  & $2.49999(5)$   &   $0.1775290(1)$ &\ \ \ \  0.1666...\ \ \ \ & $0.1128(1)$  &  0.1329807...   &  2.878(5)       \\[2pt]
  4  & $2.50000(5)$   &   $0.1340937(1)$ & 0.125 &$0.0830(1)$   &  0.0997355...   & 2.770(7)     \\[2pt]
  5  & $2.49999(5)$   &   $0.1077242(1)$ & 0.1 & $0.0657(1)$  &  0.0797884...  &  2.712(9)    \\[2pt]
  10 & $2.49999(5)$  &   $0.05430571(5)$ & 0.05 & $0.0322(1)$  &  0.0398942... &  2.61(2)  \\[2pt]
  20 & $2.50001(5)$  &   $0.02726238(2)$& 0.025 & $0.01595(5)$ & 0.0199471... &  2.56(2)   
\end{tabular}
\end{center}
\caption{ 
Values of critical exponent $\tau$, critical point $t_c$, and critical amplitudes $a_0$ and $B$ for 
different $m$ in the case of the evolution starting from isolated nodes. 
These values are obtained from the numerical solution of Eq.~(\ref{11})
 for $s\leq 10^5$ by our method \cite{da2014critical}. 
For comparison, the table shows the estimates $t_c^e=1/(2m)$ and $a_0^e=1/(\sqrt{2\pi}\, m)$ found in Sec.~\ref{s5}. The estimate for $B$ is independent of $m$, $B^e=4$.
}
\label{t1}
\end{table*}


\section{Susceptibility and order parameter}
\label{s4}

According to Ref. \cite{da2014solution}, the order parameter and generalized susceptibility for this class of problems are related to the probability $c_2$ that two nodes selected by the model rules fall within the same cluster, 
\begin{equation}
c_2= \sum_s \frac{s Q(s)^2}{N P(s)} + [1-(1-S)^m]^2\equiv \chi/N+O^2
.
\label{124}
\end{equation}
The first term on the right-hand side is the probability that both selected nodes belong to the same finite cluster, which is equal to the susceptibility $\chi$ divided by $N$. The second term is the probability that both nodes belong in the percolation cluster, which is equal to the square of the order parameter $O$. In the models under consideration, the susceptibility near the critical threshold $t_c$ 
is
\begin{equation}
\chi= \sum_s \frac{s Q(s)^2}{P(s)} \cong  m^2 \sum_s s P(s) =m^2 \langle s \rangle_P
,
\label{125}
\end{equation}
where we used Eq.~(\ref{13}), and $\langle s \rangle_P$ is the first moment of the distribution $P(s)$. For $t>t_c$, summing both sides of Eq.~(\ref{11}) over $s$ we get
\begin{equation}
\frac{\partial S}{\partial t} = 2 [1-(1-S)^m] \sum_s s Q(s) \cong 2 m^2 S  \langle s \rangle_P
.
\label{126}
\end{equation}
Similarly, for $t<t_c$, multiplying both sides of Eq.~(\ref{11}) by $s$ and summing over $s$ we obtain 
\begin{equation}
\frac{\partial \langle s \rangle_P}{\partial t} = 2 \left(\sum_s sQ(s)\right) ^2 \cong 2 m^2 \langle s \rangle_P ^2 
.
\label{127}
\end{equation}
Using  Eqs.~(\ref{123}),~(\ref{126}), and~(\ref{127}) we find that the first moment of the distribution $P(s)$ is symmetric below and above $t_c$, namely $ \langle s \rangle_P \cong (2 m^2)^{-1} |t{-}t_c|^{-1}$.    
Then, the asymptotics of the susceptibility is independent of $m$,
\begin{equation}
\chi\cong \frac{1}{2}|t-t_c|^{-1}
.
\label{128}
\end{equation}
The critical singularity of the order parameter $O = 1- (1-S)^m\cong m S$ is 
\begin{equation}
O\cong 8 \pi m^3 a_0^2 (t-t_c)
,
\label{129}
\end{equation}
where we have used Eq.~(\ref{123}). Notice that, in contrast to $\chi$, the critical amplitude of the order parameter $O$ depends on $a_0$ and $m$.



\section{Estimates}
\label{s5}

Let us estimate $P(s,t)$ by substituting the approximated relation~(\ref{13}) into the evolution equation (\ref{11}),
\begin{equation}
\frac{\partial P(s,t)}{\partial t}
\cong m^2 s \sum_{u=1}^{s-1} P(u,t)P(s{-}u,t) - 2 m sP(s,t)
.
\label{130}
\end{equation}
which is valid for large $s$ and $t$ close to $t_c$.
Let us rewrite the last equation in terms of the rescaled distribution $\tilde{P}(s,\tilde{t})\equiv m P(s,t)$ and time $\tilde{t}\equiv m t$,
\begin{equation}
\frac{\partial \tilde{P}(s,\tilde{t})}{\partial \tilde{t}}
\cong  s \sum_{u=1}^{s-1} \tilde{P}(u,\tilde{t})\tilde{P}(s{-}u,\tilde{t}) - 2  s \tilde{P}(s,\tilde{t})
.
\label{131}
\end{equation}
We assume that Eq.~(\ref{131}), being asymptotically exact near the critical point, describes approximately $P(s,t)$ in the full range of cluster sizes and time. 
This equation coincides with the exact Eq.~(\ref{11}) for ordinary percolation ($m=1$), which, for the initial condition $P(s,0)=\delta_{s,1}$, has the solution $P(s,t_c) \cong 1/(\sqrt{2\pi}) s^{-3/2}$ at the critical point $t_c=1/2$. 
This readily leads to the following estimates for $t_c$ and $a_0$ of our problem:
\begin{equation}
t_c^e=\frac{1}{2m}
,
\label{132}
\end{equation}
and
\begin{equation}
a_0^e=\frac{1}{\sqrt{2\pi}\, m}
.
\label{133}
\end{equation}
We also estimate the critical amplitude $B$ of the percolation cluster relative size, $S\cong B(t-t_c)$. Inserting $a_0^e$ into Eq.~(\ref{123}) gives $B^e=4$, independently of $m$. 
Table~\ref{t1} shows the numerical results for $t_c$, $a_0$, and $B$ 
for different $m$, and compares them with the estimates of this section. 
Notice that our simple estimate  
produces surprisingly accurate results for $m>1$. The estimate $t_c^e$ is especially good, with an error of only $5$ to $10\%$, while the estimate $a_0^e$ has a relative error about $2$ times larger.


\section{Conclusions}

In the present paper we have demonstrated that two types of the local optimization rule result in contrasting effects. The original Achlioptas rule based on selection of the smallest clusters for merging together drastically changes the critical features of continuous phase transition compared to ordinary percolation and delays the transition. Inverting this rule and selecting the largest clusters for merging, we arrive at qualitatively the same critical behavior as for ordinary percolation though with a percolation 
threshold at much earlier times, see Fig.~\ref{f2}(b). We have obtained the scaling functions and critical amplitudes for different $m$. Interestingly, the critical point and critical amplitudes obtained numerically are very close to our simple analytical estimates taking into account only 
clusters of large sizes. 
We have indicated the order parameter and susceptibility in these problems and verified the Curie--Weiss law for the susceptibility. 

In summary, we have applied the approach developed in Refs.~\cite{da2014solution,da2010explosive,da2014critical,da2015solution} to processes generated by inverting Achlioptas rule and quantitatively described the percolation transition in these models. One could also mix the two rules, original and inverse, in the same model. For instance, at each step apply one of the two rules at random.  Our results suggest that 
this is 
similar to the combination of the Achlioptas rule and the interconnection of random nodes, which leads to the usual explosive percolation effects \cite{da2011scaling,fan2012continuous,liu2012continuous,bastas2014method}. 
We based our conclusions on a set of models 
convenient for analytical treatment. We expect however that these conclusions are qualitatively valid for a much wider class of processes with inverse Achlioptas rules.

\acknowledgments 

This work was partially supported by the FET proactive
IP project MULTIPLEX 317532, the FCT project
EXPL/FIS-NAN/1275/2013, and by 
the project ``New Strategies Applied to Neuropathological Disorders'' (CENTRO-07-ST24-FEDER-002034) cofunded by QREN and EU.

\bibliographystyle{apsrev4-1}
\bibliography{refs}

\begin{thebibliography}{33}%
\makeatletter
\providecommand \@ifxundefined [1]{%
 \@ifx{#1\undefined}
}%
\providecommand \@ifnum [1]{%
 \ifnum #1\expandafter \@firstoftwo
 \else \expandafter \@secondoftwo
 \fi
}%
\providecommand \@ifx [1]{%
 \ifx #1\expandafter \@firstoftwo
 \else \expandafter \@secondoftwo
 \fi
}%
\providecommand \natexlab [1]{#1}%
\providecommand \enquote  [1]{``#1''}%
\providecommand \bibnamefont  [1]{#1}%
\providecommand \bibfnamefont [1]{#1}%
\providecommand \citenamefont [1]{#1}%
\providecommand \href@noop [0]{\@secondoftwo}%
\providecommand \href [0]{\begingroup \@sanitize@url \@href}%
\providecommand \@href[1]{\@@startlink{#1}\@@href}%
\providecommand \@@href[1]{\endgroup#1\@@endlink}%
\providecommand \@sanitize@url [0]{\catcode `\\12\catcode `\$12\catcode
  `\&12\catcode `\#12\catcode `\^12\catcode `\_12\catcode `\%12\relax}%
\providecommand \@@startlink[1]{}%
\providecommand \@@endlink[0]{}%
\providecommand \url  [0]{\begingroup\@sanitize@url \@url }%
\providecommand \@url [1]{\endgroup\@href {#1}{\urlprefix }}%
\providecommand \urlprefix  [0]{URL }%
\providecommand \Eprint [0]{\href }%
\providecommand \doibase [0]{http://dx.doi.org/}%
\providecommand \selectlanguage [0]{\@gobble}%
\providecommand \bibinfo  [0]{\@secondoftwo}%
\providecommand \bibfield  [0]{\@secondoftwo}%
\providecommand \translation [1]{[#1]}%
\providecommand \BibitemOpen [0]{}%
\providecommand \bibitemStop [0]{}%
\providecommand \bibitemNoStop [0]{.\EOS\space}%
\providecommand \EOS [0]{\spacefactor3000\relax}%
\providecommand \BibitemShut  [1]{\csname bibitem#1\endcsname}%
\let\auto@bib@innerbib\@empty
\bibitem [{\citenamefont {Achlioptas}\ \emph {et~al.}(2009)\citenamefont
  {Achlioptas}, \citenamefont {D'Souza},\ and\ \citenamefont
  {Spencer}}]{achlioptas2009explosive}%
  \BibitemOpen
  \bibfield  {author} {\bibinfo {author} {\bibfnamefont {D.}~\bibnamefont
  {Achlioptas}}, \bibinfo {author} {\bibfnamefont {R.~M.}\ \bibnamefont
  {D'Souza}}, \ and\ \bibinfo {author} {\bibfnamefont {J.}~\bibnamefont
  {Spencer}},\ }\href@noop {} {\bibfield  {journal} {\bibinfo  {journal}
  {Science}\ }\textbf {\bibinfo {volume} {323}},\ \bibinfo {pages} {1453}
  (\bibinfo {year} {2009})}\BibitemShut {NoStop}%
\bibitem [{\citenamefont {Spencer}\ and\ \citenamefont
  {Wormald}(2007)}]{spencer2007birth}%
  \BibitemOpen
  \bibfield  {author} {\bibinfo {author} {\bibfnamefont {J.}~\bibnamefont
  {Spencer}}\ and\ \bibinfo {author} {\bibfnamefont {N.}~\bibnamefont
  {Wormald}},\ }\href@noop {} {\bibfield  {journal} {\bibinfo  {journal}
  {Combinatorica}\ }\textbf {\bibinfo {volume} {27}},\ \bibinfo {pages} {587}
  (\bibinfo {year} {2007})}\BibitemShut {NoStop}%
\bibitem [{\citenamefont {Ziff}(2009)}]{ziff2009explosive}%
  \BibitemOpen
  \bibfield  {author} {\bibinfo {author} {\bibfnamefont {R.~M.}\ \bibnamefont
  {Ziff}},\ }\href@noop {} {\bibfield  {journal} {\bibinfo  {journal} {Phys.
  Rev. Lett.}\ }\textbf {\bibinfo {volume} {103}},\ \bibinfo {pages} {045701}
  (\bibinfo {year} {2009})}\BibitemShut {NoStop}%
\bibitem [{\citenamefont {da~Costa}\ \emph {et~al.}(2010)\citenamefont
  {da~Costa}, \citenamefont {Dorogovtsev}, \citenamefont {Goltsev},\ and\
  \citenamefont {Mendes}}]{da2010explosive}%
  \BibitemOpen
  \bibfield  {author} {\bibinfo {author} {\bibfnamefont {R.~A.}\ \bibnamefont
  {da~Costa}}, \bibinfo {author} {\bibfnamefont {S.~N.}\ \bibnamefont
  {Dorogovtsev}}, \bibinfo {author} {\bibfnamefont {A.~V.}\ \bibnamefont
  {Goltsev}}, \ and\ \bibinfo {author} {\bibfnamefont {J.~F.~F.}\ \bibnamefont
  {Mendes}},\ }\href@noop {} {\bibfield  {journal} {\bibinfo  {journal} {Phys.
  Rev. Lett.}\ }\textbf {\bibinfo {volume} {105}},\ \bibinfo {pages} {255701}
  (\bibinfo {year} {2010})}\BibitemShut {NoStop}%
\bibitem [{\citenamefont {D'Souza}\ and\ \citenamefont
  {Mitzenmacher}(2010)}]{d2010local}%
  \BibitemOpen
  \bibfield  {author} {\bibinfo {author} {\bibfnamefont {R.~M.}\ \bibnamefont
  {D'Souza}}\ and\ \bibinfo {author} {\bibfnamefont {M.}~\bibnamefont
  {Mitzenmacher}},\ }\href@noop {} {\bibfield  {journal} {\bibinfo  {journal}
  {Phys. Rev. Lett.}\ }\textbf {\bibinfo {volume} {104}},\ \bibinfo {pages}
  {195702} (\bibinfo {year} {2010})}\BibitemShut {NoStop}%
\bibitem [{\citenamefont {Nagler}\ \emph {et~al.}(2011)\citenamefont {Nagler},
  \citenamefont {Levina},\ and\ \citenamefont {Timme}}]{nagler2011impact}%
  \BibitemOpen
  \bibfield  {author} {\bibinfo {author} {\bibfnamefont {J.}~\bibnamefont
  {Nagler}}, \bibinfo {author} {\bibfnamefont {A.}~\bibnamefont {Levina}}, \
  and\ \bibinfo {author} {\bibfnamefont {M.}~\bibnamefont {Timme}},\
  }\href@noop {} {\bibfield  {journal} {\bibinfo  {journal} {Nature Phys.}\
  }\textbf {\bibinfo {volume} {7}},\ \bibinfo {pages} {265} (\bibinfo {year}
  {2011})}\BibitemShut {NoStop}%
\bibitem [{\citenamefont {Ara{\'u}jo}\ \emph {et~al.}(2011)\citenamefont
  {Ara{\'u}jo}, \citenamefont {Andrade}, \citenamefont {Ziff},\ and\
  \citenamefont {Herrmann}}]{araujo2011tricritical}%
  \BibitemOpen
  \bibfield  {author} {\bibinfo {author} {\bibfnamefont {N.~A.~M.}\
  \bibnamefont {Ara{\'u}jo}}, \bibinfo {author} {\bibfnamefont {J.~S.}\
  \bibnamefont {Andrade}}, \bibinfo {author} {\bibfnamefont {R.~M.}\
  \bibnamefont {Ziff}}, \ and\ \bibinfo {author} {\bibfnamefont {H.~J.}\
  \bibnamefont {Herrmann}},\ }\href@noop {} {\bibfield  {journal} {\bibinfo
  {journal} {Phys. Rev. Lett.}\ }\textbf {\bibinfo {volume} {106}},\ \bibinfo
  {pages} {095703} (\bibinfo {year} {2011})}\BibitemShut {NoStop}%
\bibitem [{\citenamefont {Grassberger}\ \emph {et~al.}(2011)\citenamefont
  {Grassberger}, \citenamefont {Christensen}, \citenamefont {Bizhani},
  \citenamefont {Son},\ and\ \citenamefont
  {Paczuski}}]{grassberger2011explosive}%
  \BibitemOpen
  \bibfield  {author} {\bibinfo {author} {\bibfnamefont {P.}~\bibnamefont
  {Grassberger}}, \bibinfo {author} {\bibfnamefont {C.}~\bibnamefont
  {Christensen}}, \bibinfo {author} {\bibfnamefont {G.}~\bibnamefont
  {Bizhani}}, \bibinfo {author} {\bibfnamefont {S.-W.}\ \bibnamefont {Son}}, \
  and\ \bibinfo {author} {\bibfnamefont {M.}~\bibnamefont {Paczuski}},\
  }\href@noop {} {\bibfield  {journal} {\bibinfo  {journal} {Phys. Rev. Lett.}\
  }\textbf {\bibinfo {volume} {106}},\ \bibinfo {pages} {225701} (\bibinfo
  {year} {2011})}\BibitemShut {NoStop}%
\bibitem [{\citenamefont {da~Costa}\ \emph
  {et~al.}(2014{\natexlab{a}})\citenamefont {da~Costa}, \citenamefont
  {Dorogovtsev}, \citenamefont {Goltsev},\ and\ \citenamefont
  {Mendes}}]{da2014critical}%
  \BibitemOpen
  \bibfield  {author} {\bibinfo {author} {\bibfnamefont {R.~A.}\ \bibnamefont
  {da~Costa}}, \bibinfo {author} {\bibfnamefont {S.~N.}\ \bibnamefont
  {Dorogovtsev}}, \bibinfo {author} {\bibfnamefont {A.~V.}\ \bibnamefont
  {Goltsev}}, \ and\ \bibinfo {author} {\bibfnamefont {J.~F.~F.}\ \bibnamefont
  {Mendes}},\ }\href@noop {} {\bibfield  {journal} {\bibinfo  {journal} {Phys.
  Rev. E}\ }\textbf {\bibinfo {volume} {89}},\ \bibinfo {pages} {042148}
  (\bibinfo {year} {2014}{\natexlab{a}})}\BibitemShut {NoStop}%
\bibitem [{\citenamefont {Fortunato}\ and\ \citenamefont
  {Radicchi}(2011)}]{fortunato2011explosive}%
  \BibitemOpen
  \bibfield  {author} {\bibinfo {author} {\bibfnamefont {S.}~\bibnamefont
  {Fortunato}}\ and\ \bibinfo {author} {\bibfnamefont {F.}~\bibnamefont
  {Radicchi}},\ }\href@noop {} {\bibfield  {journal} {\bibinfo  {journal} {J.
  Phys.: Conf. Ser.}\ }\textbf {\bibinfo {volume} {297}},\ \bibinfo {pages}
  {012009} (\bibinfo {year} {2011})}\BibitemShut {NoStop}%
\bibitem [{\citenamefont {Lee}\ \emph {et~al.}(2011)\citenamefont {Lee},
  \citenamefont {Kim},\ and\ \citenamefont {Park}}]{lee2011continuity}%
  \BibitemOpen
  \bibfield  {author} {\bibinfo {author} {\bibfnamefont {H.~K.}\ \bibnamefont
  {Lee}}, \bibinfo {author} {\bibfnamefont {B.~J.}\ \bibnamefont {Kim}}, \ and\
  \bibinfo {author} {\bibfnamefont {H.}~\bibnamefont {Park}},\ }\href@noop {}
  {\bibfield  {journal} {\bibinfo  {journal} {Phys. Rev. E}\ }\textbf {\bibinfo
  {volume} {84}},\ \bibinfo {pages} {020101} (\bibinfo {year}
  {2011})}\BibitemShut {NoStop}%
\bibitem [{\citenamefont {Riordan}\ and\ \citenamefont
  {Warnke}(2011)}]{riordan2011explosive}%
  \BibitemOpen
  \bibfield  {author} {\bibinfo {author} {\bibfnamefont {O.}~\bibnamefont
  {Riordan}}\ and\ \bibinfo {author} {\bibfnamefont {L.}~\bibnamefont
  {Warnke}},\ }\href@noop {} {\bibfield  {journal} {\bibinfo  {journal}
  {Science}\ }\textbf {\bibinfo {volume} {333}},\ \bibinfo {pages} {322}
  (\bibinfo {year} {2011})}\BibitemShut {NoStop}%
\bibitem [{\citenamefont {Riordan}\ and\ \citenamefont
  {Warnke}(2012)}]{riordan2012achlioptas}%
  \BibitemOpen
  \bibfield  {author} {\bibinfo {author} {\bibfnamefont {O.}~\bibnamefont
  {Riordan}}\ and\ \bibinfo {author} {\bibfnamefont {L.}~\bibnamefont
  {Warnke}},\ }\href@noop {} {\bibfield  {journal} {\bibinfo  {journal} {Ann.
  Appl. Probab.}\ }\textbf {\bibinfo {volume} {22}},\ \bibinfo {pages} {1450}
  (\bibinfo {year} {2012})}\BibitemShut {NoStop}%
\bibitem [{\citenamefont {Cho}\ \emph {et~al.}(2013)\citenamefont {Cho},
  \citenamefont {Hwang}, \citenamefont {Herrmann},\ and\ \citenamefont
  {Kahng}}]{cho2013avoiding}%
  \BibitemOpen
  \bibfield  {author} {\bibinfo {author} {\bibfnamefont {Y.~S.}\ \bibnamefont
  {Cho}}, \bibinfo {author} {\bibfnamefont {S.}~\bibnamefont {Hwang}}, \bibinfo
  {author} {\bibfnamefont {H.~J.}\ \bibnamefont {Herrmann}}, \ and\ \bibinfo
  {author} {\bibfnamefont {B.}~\bibnamefont {Kahng}},\ }\href@noop {}
  {\bibfield  {journal} {\bibinfo  {journal} {Science}\ }\textbf {\bibinfo
  {volume} {339}},\ \bibinfo {pages} {1185} (\bibinfo {year}
  {2013})}\BibitemShut {NoStop}%
\bibitem [{\citenamefont {Ara{\'u}jo}\ \emph {et~al.}(2014)\citenamefont
  {Ara{\'u}jo}, \citenamefont {Grassberger}, \citenamefont {Kahng},
  \citenamefont {Schrenk},\ and\ \citenamefont {Ziff}}]{araujo2014recent}%
  \BibitemOpen
  \bibfield  {author} {\bibinfo {author} {\bibfnamefont {N.}~\bibnamefont
  {Ara{\'u}jo}}, \bibinfo {author} {\bibfnamefont {P.}~\bibnamefont
  {Grassberger}}, \bibinfo {author} {\bibfnamefont {B.}~\bibnamefont {Kahng}},
  \bibinfo {author} {\bibfnamefont {K.}~\bibnamefont {Schrenk}}, \ and\
  \bibinfo {author} {\bibfnamefont {R.}~\bibnamefont {Ziff}},\ }\href@noop {}
  {\bibfield  {journal} {\bibinfo  {journal} {Eur. Phys. J. ST}\ }\textbf
  {\bibinfo {volume} {223}},\ \bibinfo {pages} {2307} (\bibinfo {year}
  {2014})}\BibitemShut {NoStop}%
\bibitem [{\citenamefont {Bastas}\ \emph
  {et~al.}(2014{\natexlab{a}})\citenamefont {Bastas}, \citenamefont
  {Giazitzidis}, \citenamefont {Maragakis},\ and\ \citenamefont
  {Kosmidis}}]{bastas2014explosive}%
  \BibitemOpen
  \bibfield  {author} {\bibinfo {author} {\bibfnamefont {N.}~\bibnamefont
  {Bastas}}, \bibinfo {author} {\bibfnamefont {P.}~\bibnamefont {Giazitzidis}},
  \bibinfo {author} {\bibfnamefont {M.}~\bibnamefont {Maragakis}}, \ and\
  \bibinfo {author} {\bibfnamefont {K.}~\bibnamefont {Kosmidis}},\ }\href@noop
  {} {\bibfield  {journal} {\bibinfo  {journal} {Physica A}\ }\textbf {\bibinfo
  {volume} {407}},\ \bibinfo {pages} {54} (\bibinfo {year}
  {2014}{\natexlab{a}})}\BibitemShut {NoStop}%
\bibitem [{\citenamefont {da~Costa}\ \emph
  {et~al.}(2014{\natexlab{b}})\citenamefont {da~Costa}, \citenamefont
  {Dorogovtsev}, \citenamefont {Goltsev},\ and\ \citenamefont
  {Mendes}}]{da2014solution}%
  \BibitemOpen
  \bibfield  {author} {\bibinfo {author} {\bibfnamefont {R.~A.}\ \bibnamefont
  {da~Costa}}, \bibinfo {author} {\bibfnamefont {S.~N.}\ \bibnamefont
  {Dorogovtsev}}, \bibinfo {author} {\bibfnamefont {A.~V.}\ \bibnamefont
  {Goltsev}}, \ and\ \bibinfo {author} {\bibfnamefont {J.~F.~F.}\ \bibnamefont
  {Mendes}},\ }\href@noop {} {\bibfield  {journal} {\bibinfo  {journal} {Phys.
  Rev. E}\ }\textbf {\bibinfo {volume} {90}},\ \bibinfo {pages} {022145}
  (\bibinfo {year} {2014}{\natexlab{b}})}\BibitemShut {NoStop}%
\bibitem [{\citenamefont {Stauffer}\ and\ \citenamefont
  {Aharony}(1991)}]{stauffer1991introduction}%
  \BibitemOpen
  \bibfield  {author} {\bibinfo {author} {\bibfnamefont {D.}~\bibnamefont
  {Stauffer}}\ and\ \bibinfo {author} {\bibfnamefont {A.}~\bibnamefont
  {Aharony}},\ }\href@noop {} {\emph {\bibinfo {title} {{Introduction to
  Percolation Theory}}}}\ (\bibinfo  {publisher} {Taylor and Francis},\
  \bibinfo {address} {London},\ \bibinfo {year} {1991})\BibitemShut {NoStop}%
\bibitem [{\citenamefont {Stauffer}(1979)}]{stauffer1979scaling}%
  \BibitemOpen
  \bibfield  {author} {\bibinfo {author} {\bibfnamefont {D.}~\bibnamefont
  {Stauffer}},\ }\href@noop {} {\bibfield  {journal} {\bibinfo  {journal}
  {Phys. Rep.}\ }\textbf {\bibinfo {volume} {54}},\ \bibinfo {pages} {1}
  (\bibinfo {year} {1979})}\BibitemShut {NoStop}%
\bibitem [{\citenamefont {Dorogovtsev}\ \emph {et~al.}(2008)\citenamefont
  {Dorogovtsev}, \citenamefont {Goltsev},\ and\ \citenamefont
  {Mendes}}]{dorogovtsev2008critical}%
  \BibitemOpen
  \bibfield  {author} {\bibinfo {author} {\bibfnamefont {S.~N.}\ \bibnamefont
  {Dorogovtsev}}, \bibinfo {author} {\bibfnamefont {A.~V.}\ \bibnamefont
  {Goltsev}}, \ and\ \bibinfo {author} {\bibfnamefont {J.~F.~F.}\ \bibnamefont
  {Mendes}},\ }\href@noop {} {\bibfield  {journal} {\bibinfo  {journal} {Rev.
  Mod. Phys.}\ }\textbf {\bibinfo {volume} {80}},\ \bibinfo {pages} {1275}
  (\bibinfo {year} {2008})}\BibitemShut {NoStop}%
\bibitem [{\citenamefont {Rozenfeld}\ \emph {et~al.}(2010)\citenamefont
  {Rozenfeld}, \citenamefont {Gallos},\ and\ \citenamefont
  {Makse}}]{rozenfeld2010explosive}%
  \BibitemOpen
  \bibfield  {author} {\bibinfo {author} {\bibfnamefont {H.~D.}\ \bibnamefont
  {Rozenfeld}}, \bibinfo {author} {\bibfnamefont {L.~K.}\ \bibnamefont
  {Gallos}}, \ and\ \bibinfo {author} {\bibfnamefont {H.~A.}\ \bibnamefont
  {Makse}},\ }\href@noop {} {\bibfield  {journal} {\bibinfo  {journal} {Eur.
  Phys. J. B}\ }\textbf {\bibinfo {volume} {75}},\ \bibinfo {pages} {305}
  (\bibinfo {year} {2010})}\BibitemShut {NoStop}%
\bibitem [{\citenamefont {Kim}\ \emph {et~al.}(2010)\citenamefont {Kim},
  \citenamefont {Yun},\ and\ \citenamefont {Yook}}]{kim2010explosive}%
  \BibitemOpen
  \bibfield  {author} {\bibinfo {author} {\bibfnamefont {Y.}~\bibnamefont
  {Kim}}, \bibinfo {author} {\bibfnamefont {Y.-k.}\ \bibnamefont {Yun}}, \ and\
  \bibinfo {author} {\bibfnamefont {S.-H.}\ \bibnamefont {Yook}},\ }\href@noop
  {} {\bibfield  {journal} {\bibinfo  {journal} {Phys. Rev. E}\ }\textbf
  {\bibinfo {volume} {82}},\ \bibinfo {pages} {061105} (\bibinfo {year}
  {2010})}\BibitemShut {NoStop}%
\bibitem [{\citenamefont {Pan}\ \emph {et~al.}(2011)\citenamefont {Pan},
  \citenamefont {Kivel{\"a}}, \citenamefont {Saram{\"a}ki}, \citenamefont
  {Kaski},\ and\ \citenamefont {Kert{\'e}sz}}]{pan2011using}%
  \BibitemOpen
  \bibfield  {author} {\bibinfo {author} {\bibfnamefont {R.~K.}\ \bibnamefont
  {Pan}}, \bibinfo {author} {\bibfnamefont {M.}~\bibnamefont {Kivel{\"a}}},
  \bibinfo {author} {\bibfnamefont {J.}~\bibnamefont {Saram{\"a}ki}}, \bibinfo
  {author} {\bibfnamefont {K.}~\bibnamefont {Kaski}}, \ and\ \bibinfo {author}
  {\bibfnamefont {J.}~\bibnamefont {Kert{\'e}sz}},\ }\href@noop {} {\bibfield
  {journal} {\bibinfo  {journal} {Phys. Rev. E}\ }\textbf {\bibinfo {volume}
  {83}},\ \bibinfo {pages} {046112} (\bibinfo {year} {2011})}\BibitemShut
  {NoStop}%
\bibitem [{\citenamefont {D'Souza}\ \emph {et~al.}(2007)\citenamefont
  {D'Souza}, \citenamefont {Krapivsky},\ and\ \citenamefont
  {Moore}}]{d2007power}%
  \BibitemOpen
  \bibfield  {author} {\bibinfo {author} {\bibfnamefont {R.~M.}\ \bibnamefont
  {D'Souza}}, \bibinfo {author} {\bibfnamefont {P.~L.}\ \bibnamefont
  {Krapivsky}}, \ and\ \bibinfo {author} {\bibfnamefont {C.}~\bibnamefont
  {Moore}},\ }\href@noop {} {\bibfield  {journal} {\bibinfo  {journal} {Eur.
  Phys. J. B}\ }\textbf {\bibinfo {volume} {59}},\ \bibinfo {pages} {535}
  (\bibinfo {year} {2007})}\BibitemShut {NoStop}%
\bibitem [{\citenamefont {Tanaka}\ and\ \citenamefont
  {Tamura}(2013)}]{tanaka2013network}%
  \BibitemOpen
  \bibfield  {author} {\bibinfo {author} {\bibfnamefont {S.}~\bibnamefont
  {Tanaka}}\ and\ \bibinfo {author} {\bibfnamefont {R.}~\bibnamefont
  {Tamura}},\ }\href@noop {} {\bibfield  {journal} {\bibinfo  {journal} {J.
  Phys. Soc. Jpn.}\ }\textbf {\bibinfo {volume} {82}},\ \bibinfo {pages}
  {053002} (\bibinfo {year} {2013})}\BibitemShut {NoStop}%
\bibitem [{\citenamefont {Giazitzidis}\ \emph {et~al.}(2014)\citenamefont
  {Giazitzidis}, \citenamefont {Avramov},\ and\ \citenamefont
  {Argyrakis}}]{giazitzidis2014variation}%
  \BibitemOpen
  \bibfield  {author} {\bibinfo {author} {\bibfnamefont {P.}~\bibnamefont
  {Giazitzidis}}, \bibinfo {author} {\bibfnamefont {I.}~\bibnamefont
  {Avramov}}, \ and\ \bibinfo {author} {\bibfnamefont {P.}~\bibnamefont
  {Argyrakis}},\ }\href@noop {} {\bibfield  {journal} {\bibinfo  {journal}
  {arXiv preprint arXiv:1411.3839}\ } (\bibinfo {year} {2014})}\BibitemShut
  {NoStop}%
\bibitem [{\citenamefont {da~Costa}\ \emph {et~al.}(2015)\citenamefont
  {da~Costa}, \citenamefont {Dorogovtsev}, \citenamefont {Goltsev},\ and\
  \citenamefont {Mendes}}]{da2015solution}%
  \BibitemOpen
  \bibfield  {author} {\bibinfo {author} {\bibfnamefont {R.~A.}\ \bibnamefont
  {da~Costa}}, \bibinfo {author} {\bibfnamefont {S.~N.}\ \bibnamefont
  {Dorogovtsev}}, \bibinfo {author} {\bibfnamefont {A.~V.}\ \bibnamefont
  {Goltsev}}, \ and\ \bibinfo {author} {\bibfnamefont {J.~F.~F.}\ \bibnamefont
  {Mendes}},\ }\href@noop {} {\bibfield  {journal} {\bibinfo  {journal} {Phys.
  Rev. E}\ }\textbf {\bibinfo {volume} {91}},\ \bibinfo {pages} {032140}
  (\bibinfo {year} {2015})}\BibitemShut {NoStop}%
\bibitem [{\citenamefont {Smoluchowski}(1916)}]{smoluchowski1916brownsche}%
  \BibitemOpen
  \bibfield  {author} {\bibinfo {author} {\bibfnamefont {M.}~\bibnamefont
  {Smoluchowski}},\ }\href@noop {} {\bibfield  {journal} {\bibinfo  {journal}
  {Annalen der Physik}\ }\textbf {\bibinfo {volume} {353}},\ \bibinfo {pages}
  {1103} (\bibinfo {year} {1916})}\BibitemShut {NoStop}%
\bibitem [{\citenamefont {Krapivsky}\ \emph {et~al.}(2010)\citenamefont
  {Krapivsky}, \citenamefont {Redner},\ and\ \citenamefont
  {Ben-Naim}}]{krapivsky2010kinetic}%
  \BibitemOpen
  \bibfield  {author} {\bibinfo {author} {\bibfnamefont {P.~L.}\ \bibnamefont
  {Krapivsky}}, \bibinfo {author} {\bibfnamefont {S.}~\bibnamefont {Redner}}, \
  and\ \bibinfo {author} {\bibfnamefont {E.}~\bibnamefont {Ben-Naim}},\
  }\href@noop {} {\emph {\bibinfo {title} {{A Kinetic View of Statistical
  Physics}}}}\ (\bibinfo  {publisher} {Cambridge University Press},\ \bibinfo
  {address} {Cambridge},\ \bibinfo {year} {2010})\BibitemShut {NoStop}%
\bibitem [{\citenamefont {da~Costa}\ \emph {et~al.}(2011)\citenamefont
  {da~Costa}, \citenamefont {Dorogovtsev}, \citenamefont {Goltsev},\ and\
  \citenamefont {Mendes}}]{da2011scaling}%
  \BibitemOpen
  \bibfield  {author} {\bibinfo {author} {\bibfnamefont {R.~A.}\ \bibnamefont
  {da~Costa}}, \bibinfo {author} {\bibfnamefont {S.~N.}\ \bibnamefont
  {Dorogovtsev}}, \bibinfo {author} {\bibfnamefont {A.~V.}\ \bibnamefont
  {Goltsev}}, \ and\ \bibinfo {author} {\bibfnamefont {J.~F.~F.}\ \bibnamefont
  {Mendes}},\ }\href@noop {} {\bibfield  {journal} {\bibinfo  {journal} {Int.
  J. Complex Systems in Science}\ }\textbf {\bibinfo {volume} {1}},\ \bibinfo
  {pages} {169} (\bibinfo {year} {2011})}\BibitemShut {NoStop}%
\bibitem [{\citenamefont {Fan}\ \emph {et~al.}(2012)\citenamefont {Fan},
  \citenamefont {Liu}, \citenamefont {Li},\ and\ \citenamefont
  {Chen}}]{fan2012continuous}%
  \BibitemOpen
  \bibfield  {author} {\bibinfo {author} {\bibfnamefont {J.}~\bibnamefont
  {Fan}}, \bibinfo {author} {\bibfnamefont {M.}~\bibnamefont {Liu}}, \bibinfo
  {author} {\bibfnamefont {L.}~\bibnamefont {Li}}, \ and\ \bibinfo {author}
  {\bibfnamefont {X.}~\bibnamefont {Chen}},\ }\href@noop {} {\bibfield
  {journal} {\bibinfo  {journal} {Phys. Rev. E}\ }\textbf {\bibinfo {volume}
  {85}},\ \bibinfo {pages} {061110} (\bibinfo {year} {2012})}\BibitemShut
  {NoStop}%
\bibitem [{\citenamefont {Liu}\ \emph {et~al.}(2012)\citenamefont {Liu},
  \citenamefont {Fan}, \citenamefont {Li},\ and\ \citenamefont
  {Chen}}]{liu2012continuous}%
  \BibitemOpen
  \bibfield  {author} {\bibinfo {author} {\bibfnamefont {M.}~\bibnamefont
  {Liu}}, \bibinfo {author} {\bibfnamefont {J.}~\bibnamefont {Fan}}, \bibinfo
  {author} {\bibfnamefont {L.}~\bibnamefont {Li}}, \ and\ \bibinfo {author}
  {\bibfnamefont {X.}~\bibnamefont {Chen}},\ }\href@noop {} {\bibfield
  {journal} {\bibinfo  {journal} {Eur. Phys. J. B}\ }\textbf {\bibinfo {volume}
  {85}},\ \bibinfo {pages} {132} (\bibinfo {year} {2012})}\BibitemShut
  {NoStop}%
\bibitem [{\citenamefont {Bastas}\ \emph
  {et~al.}(2014{\natexlab{b}})\citenamefont {Bastas}, \citenamefont {Kosmidis},
  \citenamefont {Giazitzidis},\ and\ \citenamefont
  {Maragakis}}]{bastas2014method}%
  \BibitemOpen
  \bibfield  {author} {\bibinfo {author} {\bibfnamefont {N.}~\bibnamefont
  {Bastas}}, \bibinfo {author} {\bibfnamefont {K.}~\bibnamefont {Kosmidis}},
  \bibinfo {author} {\bibfnamefont {P.}~\bibnamefont {Giazitzidis}}, \ and\
  \bibinfo {author} {\bibfnamefont {M.}~\bibnamefont {Maragakis}},\ }\href@noop
  {} {\bibfield  {journal} {\bibinfo  {journal} {Phys. Rev. E}\ }\textbf
  {\bibinfo {volume} {90}},\ \bibinfo {pages} {062101} (\bibinfo {year}
  {2014}{\natexlab{b}})}\BibitemShut {NoStop}%
\end{thebibliography}%

\end{document}